# Effect of Initial Volume on the Run-Out Behavior of Submerged Granular Columns

Qiuyu Wang, M.S.,[1] Reihaneh Hosseini, M.S.,[2] and Krishna Kumar, Ph.D., Aff. M. ASCE[3]

[1] Department of Civil, Architectural and Environmental Engineering, The University of Texas at Austin, Austin, TX 78712; email: wangqiuyu@utexas.edu
[2] Department of Civil, Architectural and Environmental Engineering, The University of Texas at Austin, Austin, TX 78712; email: reihos@utexas.edu
[3] Department of Civil, Architectural and Environmental Engineering, The University of Texas at Austin, Austin, TX 78712; email: krishnak@mail.utexas.edu

## ABSTRACT

Submarine landslides transport thousands of cubic meters of sediment across continental shelves even at slopes as low as 1° and can cause significant casualty and damage to infrastructure. The run-out mechanism in a submarine landslide is affected by factors such as the initial packing density, permeability, slope angle, and initial volume. While past studies have focused on the influence of density, permeability, and slope angle on the granular column collapse, the impact of volume on the run-out characteristics has not been investigated. This study aims to understand how the initial volume affects the run-out using a two-dimensional coupled lattice Boltzman and discrete element (LBM-DEM) method. The coupled LBM-DEM approach allows simulating fluid flow at the pore-scale resolution to understand the grain-scale mechanisms driving the complex continuum-scale response in the granular column collapse. For submerged granular column collapse, the run-out mechanism is heavily influenced by the interaction between the grains and the surrounding fluid. The development of negative pore pressures during shearing and hydrodynamic drag forces inhibit the flow. On the other hand, entrainment of water resulting in hydroplaning enhances the flow. With an increase in volume, the interaction between the grains and the surrounding fluid varies, causing changes in the run-out behavior. For smaller volumes, the forces inhibiting the underwater flow predominates, resulting in shorter run-outs than their dry counterparts. At large volumes, hydroplaning results in larger run-out than the dry cases, despite the inhibiting effects of drag forces and negative pore pressures.

## INTRODUCTION

Submerged landslides are characterized by mobilizing substantial volumes of sediments, flowing hundreds of kilometers (Korup et al. 2007). To study the complex flow dynamics of landslides researchers have often used a simple model of a collapsing granular column. While for a dry granular collapse, the final run-out distance is primarily a function of the column's initial aspect ratio (Thompson & Hupper 2007; Lajeunesse et al. 2005; Lube et al. 2005; Lube et al. 2004), for an underwater granular column collapse, the run-out distance is influenced by a number of additional factors as a result of the hydrodynamic interactions. Some of these factors have been studied previously, including initial volume fraction, permeability and geometry such as the slope angle (Yang et al. 2019; Bougouin & Lacaze 2018; Kumar et al. 2017; Pailha et al. 2013; Topin et al. 2012; Rondon et al. 2011). Loose packings flow rapidly and give rise to more elongated final deposits than the dense counterpart. The final runout also decreases with permeability, especially for loose packings. The slope geometry controls the growth of the basal shear band and thus run-out behavior (Puzrin et al. 2015, 2016).

However, the effect of volume on the run-out behavior has been seldom investigated (Staron & Lajeunesse 2009; Yang et al. 2020). Since the sizes of granular columns studied are considerably smaller, by several orders of magnitude in volume than natural landslides, understanding the effect of volume on underwater run-out behavior is crucial for reliably predicting the run-out distance for realistic volumes.

This study investigates the effect of volume on underwater run-out characteristics by numerically modeling two-dimensional (2D) submerged granular columns with varying initial volumes at a constant aspect ratio. In order to accurately capture the hydrodynamic interactions at the grain and pore-scale, a coupled lattice Boltzman and discrete element method (LBM-DEM) approach is used. The utilization of LBM-DEM allows the identification of hydrodynamic mechanisms involved in the collapse and thereby a more fundamental understanding of the influence of volume. In what follows, after a brief description of the numerical method and the simulation setup, and an overview of the collapse evolution, the role of volume on the run-out behavior is discussed in detail by looking into the effects of negative pore pressure generation, hydroplaning, drag force and formation of turbulent vortices, at different volumes and collapse stages.

**NUMERICAL METHOD**

The LBM models the fluid flow at the mesoscopic scale, making it suitable for studying pore-scale fluid behavior between soil grains. The DEM controls the interaction of individual grains involved in the column collapse. These methods are coupled by accounting for the exchange of momentum between the fluid (LBM) and the grains (DEM). For further information on the coupling between the LBM and DEM, refer to Kumar et al. (2017). An important consideration when simulating a porous media flow in 2D is the presence of non-interconnected pore space when circular grains come in contact on a plane. This is undesirable since, in reality (3D), the pore spaces between the grains are likely to allow fluid flow perpendicular to the plane of reference. This 2D artifact is resolved by adopting a hydrodynamic radius, a reduced radius of $0.8R$ only for the LBM computations (Boutt et al. 2007). The hydrodynamic radius can be thought of as a 2D representation of the 3D permeability of the granular material. The DEM computations, which rely on the grain contacts, are carried out with the actual radius of the grains.

**SIMULATION SETUP**

In this study, we investigate the effect of volume on the run-out behaviour of granular columns with an aspect ratio, defined as the ratio of the initial height ($H_i$) to the initial length ($L_i$), of 0.2. Six different initial volumes varying from 10,000 to 60,000 cm$^3$ are used to model the run-out response. The columns are constrained by a wall on one side and a gate on the other, as shown in Figure 1. The initial configuration is created by ballistic deposition of grains under dry conditions, resulting in an initial packing density of 84%. The grains are modeled as discs with a density of 2650 kg/m$^3$, contact friction angle of 28°, coefficient of restitution of 0.26, and linear spring stiffness of $1.6 \times 10^6$ N/m. The sample is polydisperse with a minimum and maximum of 0.5 and 0.9 mm, respectively. After the column reaches equilibrium, the column is placed underwater, and the gate is instantaneously removed to cause the collapse. The final run-out distance ($L_f$) and the collapsed height ($H_f$) are measured.

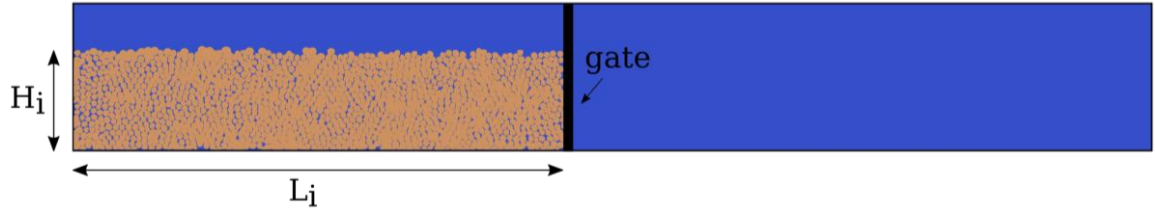

Figure 1. Configuration of underwater granular column collapse simulations

## EVOLUTION OF A GRANULAR COLLAPSE

Figure 2 shows the snapshots of the run-out evolution of a granular column with an initial aspect ratio of 0.2 and a volume of 60,000 cm$^3$. The evolution of a granular column collapse can be classified into three stages: (a) initiation, (b) spreading, and (c) settlement. The initiation stage is characterized by the mobilization of grains above the failure plane inclined at an angle of 45-50° from the base of the column (Figure 2a - b). In the dry case, Lajeunesse et al. (2005) observed that the grains above the failure plane are fully mobilized at a characteristic time $\tau_c = \sqrt{(H/g)}$. In the presence of fluid, the initiation stage is prolonged due to the hydrodynamic interactions and lasts until $3\tau_c$. The run-out spreading stage ($3\tau_c < t < 6\tau_c$) involves the translation of the initial potential energy to horizontal spreading. In submerged cases, this stage is also accompanied by the distinct development of turbulent eddies (Figure 2c-d). Lastly, the settlement phase ($t > 6\tau_c$) is characterized by the slow down and settlement of the sliding granular mass, which results in an increase in the packing density. As the granular mass loses horizontal velocity and settles, the eddies developed in the horizontal spreading stage start to depart from the free surface (Figure 2e).

## THE ROLE OF INITIAL VOLUME

The effect of initial volume on the run-out behavior is demonstrated in Figure 3 by plotting the normalized run-out with respect to normalized time for dry and submerged columns with different initial volumes (10,000 to 60,000 cm$^3$). The normalized run-out distance is defined as ($L_f - L_i)/L_i$, where the final run-out distance ($L_f$) is measured at the furthest grain with at least three contacts to the main mass to avoid a runaway grain affecting the run-out distance. The following observations are made: 1. In both dry and submerged cases, the final normalized run-out increases with volume. 2. For the smallest volume (10,000 cm$^3$), the normalized final run-out distance of the submerged column is 10% lower than the dry column, while for the largest volume (60,000 cm$^3$), the submerged column flows about 5% further than the dry column. 3. The rate of run-out evolution (slope of the curves in Figure 3a) is faster for the dry columns as compared to submerged columns. 4. For the submerged columns, the rate of run-out at initiation (t < 3 $\tau_c$) increases with volume.

Figure 4(a) shows the evolution of kinetic energy, normalized with respect to the initial potential energy ($E_0$) versus normalized time for dry and submerged columns with different initial volumes. Note that the dry collapse comes to rest at $6\tau_c$ and the normalized kinetic energy above $6\tau_c$ includes spurious oscillations due to a few individual run-away grains; therefore, it is not shown in Figure 4(a). In addition, Figure 4(b) shows the total normalized kinetic energy (area under each curve in Figure 4(a)) for different volumes in dry and submerged cases. It can be observed from Figure 4 that: 1. The total normalized kinetic energy increases with volume for both dry and submerged cases. 2. The peak kinetic energy for underwater cases is 30-40% lower than the dry cases. Also, the total normalized kinetic energy is less for submerged cases than the dry cases, and this difference decreases as the volume increases. 3. The occurrence of the peak kinetic energy shifts from $\tau_c$ in the dry cases to $3\tau_c$ in submerged cases, indicating slower run-out evolution in the submerged cases. 4. For the submerged columns, the normalized kinetic energy at initiation (t < 3 $\tau_c$) increases with volume. In the following sections, different mechanisms that contribute to the observations made above are discussed.

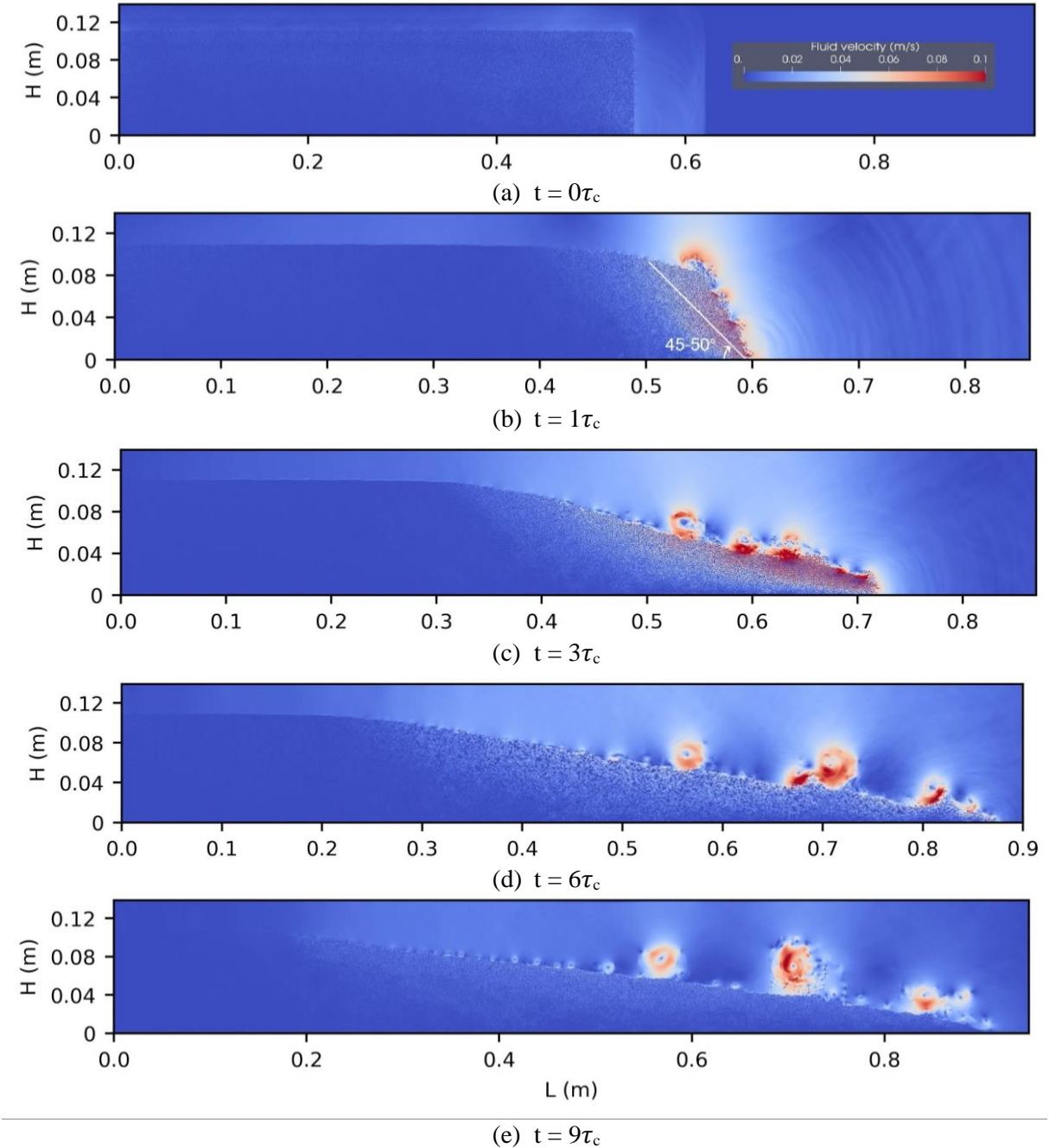

(a) t = $0\tau_c$

(b) t = $1\tau_c$

(c) t = $3\tau_c$

(d) t = $6\tau_c$

(e) t = $9\tau_c$

**Figure 2. Flow evolution of a granular column collapse in fluid (a = 0.2, V = 60,000 cm³)**

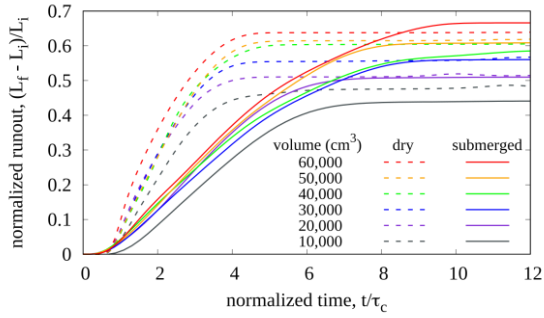
(a) Normalized runout evolution

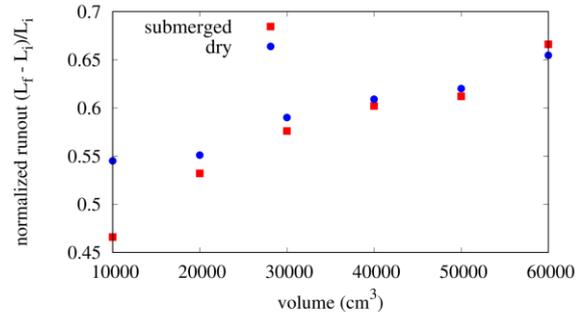
(b) Final normalized runout

**Figure 3. The normalized run-out distance (based on the furthest grain with at least 3 contacts to the main mass) in the dry and submerged cases for varying initial volumes**

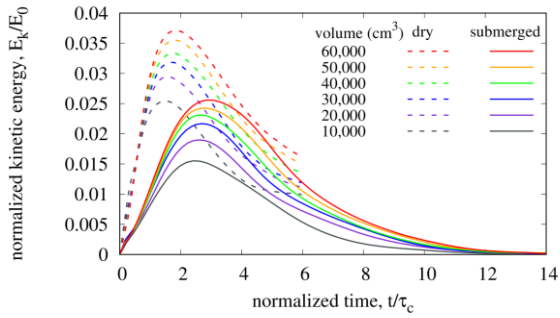
(c) Normalized kinetic energy

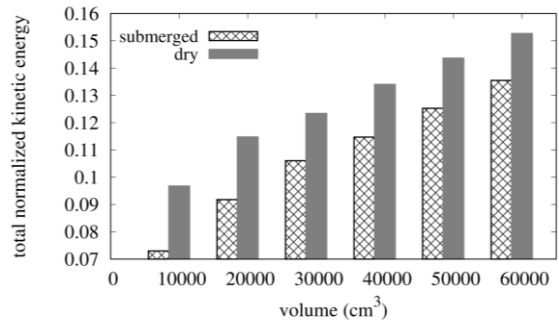
(d) Total normalized kinetic energy

**Figure 4. Kinetic energy for different volumes in dry and submerged cases**

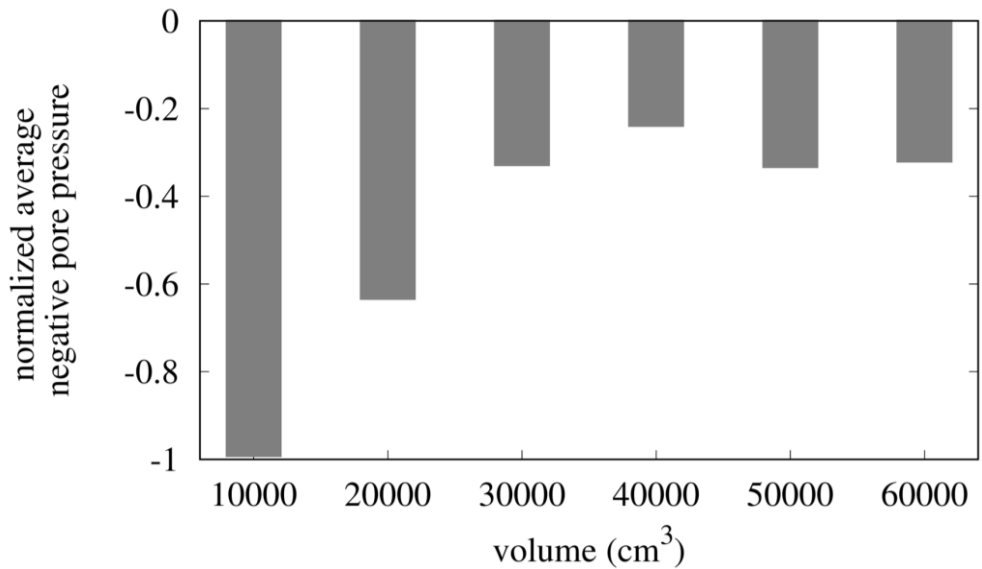

**Figure 5. normalized average pore pressures along the failure plane at collapse initiation for different volumes at $\tau_c$**

**The effect of negative pore pressure generation.** During the initiation stage of a submerged granular collapse (t < 3 $\tau_c$), as dense granular material is sheared along the failure surface, negative pore pressures develop. The mobilization of granular mass above the failure surface requires overcoming this large negative pore pressure, resulting in a slower run-out evolution than dry cases. Figure 5 shows the measured negative excess pore pressures along the shearing plane at $\tau_c$, for different initial volumes. The pore pressure is normalized by the initial maximum pore pressure $\gamma_\omega H_i$ and averaged over the length of the shearing plane. For smaller volumes, a larger amount of negative pore pressure develops and affects a higher proportion of the sliding mass, resulting in a slower run-out.

**The influence of water entrainment.** During the run-out spreading stage, energy is lost due to frictional dissipation at the base. However, in the submerged cases, this frictional dissipation is reduced due to water entrainment as observed by the decrease in the effective stresses with the volume. The loss of frictional resistance due to entrainment of water at the flow front is called hydroplaning. Harbitz (2003) observed hydroplaning of submerged flows above a critical value of densimetric Froude's number of 0.4. Froude's number is defined as the ratio of flow inertia to gravity and is expressed as $F_{rd} = U / \sqrt{\frac{\rho_d}{\rho_\omega} gH}$ where U is the average velocity of the sliding mass at flow front, $\rho_d$ and $\rho_\omega$ are the density of soil and water, respectively, H is the average thickness of the flow front, and $g$ is the gravitational acceleration. Figure 6 shows the evolution of Froude's number for different volumes. For all the volumes except 10,000 cm³, the Froude's number is larger than 0.4, which indicates the potential for hydroplaning. Once the hydroplaning occurs, the movement of flow fronts is facilitated by a significant reduction in basal friction. In addition to the amount of hydroplaning, the duration of hydroplaning also increases with an increase in volume, thereby enabling longer run-out distances. For the volume of 60,000 cm³, the peak value of Froude's number reaches 0.8, and the duration of hydroplaning also lasts a total of 6 $\tau_c$, compared to 20,000 cm³ where the peak is about 0.57 and the duration is only 3$\tau_c$. Referring to observation 2, although the amount of total normalized kinetic energy in the submerged cases is smaller than the dry cases (-13% for the largest volume based on Figure 4(b)) as a result of drag forces and turbulent vortices, due to the high amount and long duration of hydroplaning for volumes greater than 10,000 cm³, all volumes except 10,000 cm³ exhibit an almost equal or larger final run-out distance than their dry counterparts (+5% for the largest volume based on Figure 3).

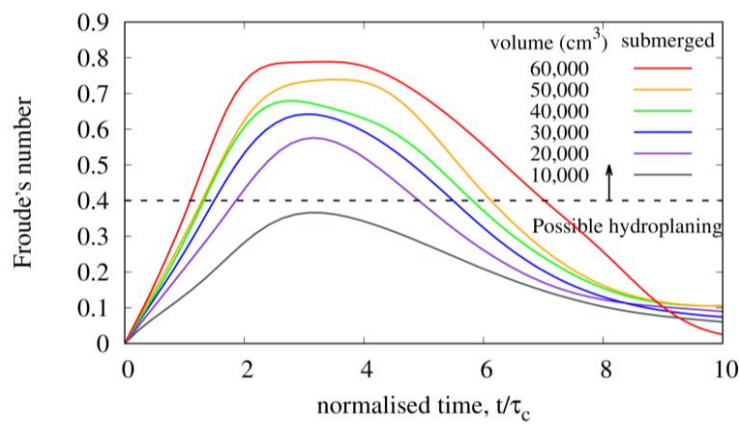

**Figure 6. The evolution of Froude's number with time for different volumes**

For submerged volumes greater than 10,000 cm$^3$, hydroplaning results in an almost equal or larger final run-out distance as their dry counterparts (+5% for the largest volume), even though the amount of total normalized kinetic energy in the submerged cases is smaller than the dry cases (-13% for the largest volume) as a result of drag forces and turbulent vortices discussed next.

**The effect of drag forces and turbulent vortices.** During the run-out spreading and settlement stage of the submerged collapse (t > 3 $\tau_c$), energy is dissipated due to the drag forces at the flow front and formation of turbulent vortices on the granular surface. The effect of the drag forces and turbulent vortices are investigated by looking into the flow morphology (Figure 7 and 8), the angle of attack of the granular collapse (Table 1), and the streamlines at the flow front (Figure 9). Figure 7 shows an irregular curved morphology in the submerged case at t > 3 $\tau_c$. Figure 8 demonstrates that as the volume increases, the granular surface interacting with the surrounding fluid as well as the number of turbulent vortices formed near the flowing granular surface increase. At 10,000 cm$^3$, only one eddy is observed at the flow front, with a local fluid velocity inside the eddy reaching 0.6 m/s. In contrast, the largest volume has at least three eddies developing at the flow front, whirling the grains away from the surface. Table 1 shows that, at any volume, the angle of attack in the submerged case is higher than the dry case. In the submerged case the angle of attack increases with volume, while in the dry case it remains roughly constant. The higher angle of attack in the submerged cases is due to the drag forces at the flow front, as shown by the increase in the streamlines at the flow front in Figure 8, which reduces the velocity of the grains creating steeper flow front. The drag forces and the formation of turbulent vortices result in a lower amount of total normalized kinetic energy for the submerged cases as compared to the dry cases.

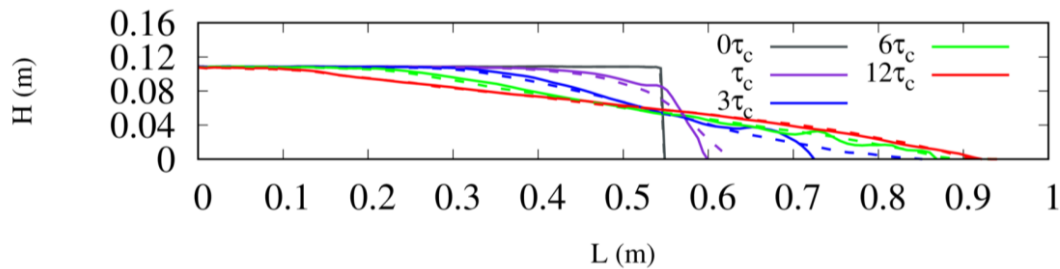

**Figure 7. Profile outline evolution for dry (dashed line) and submerged (solid line) cases for V = 60,000 cm$^3$**

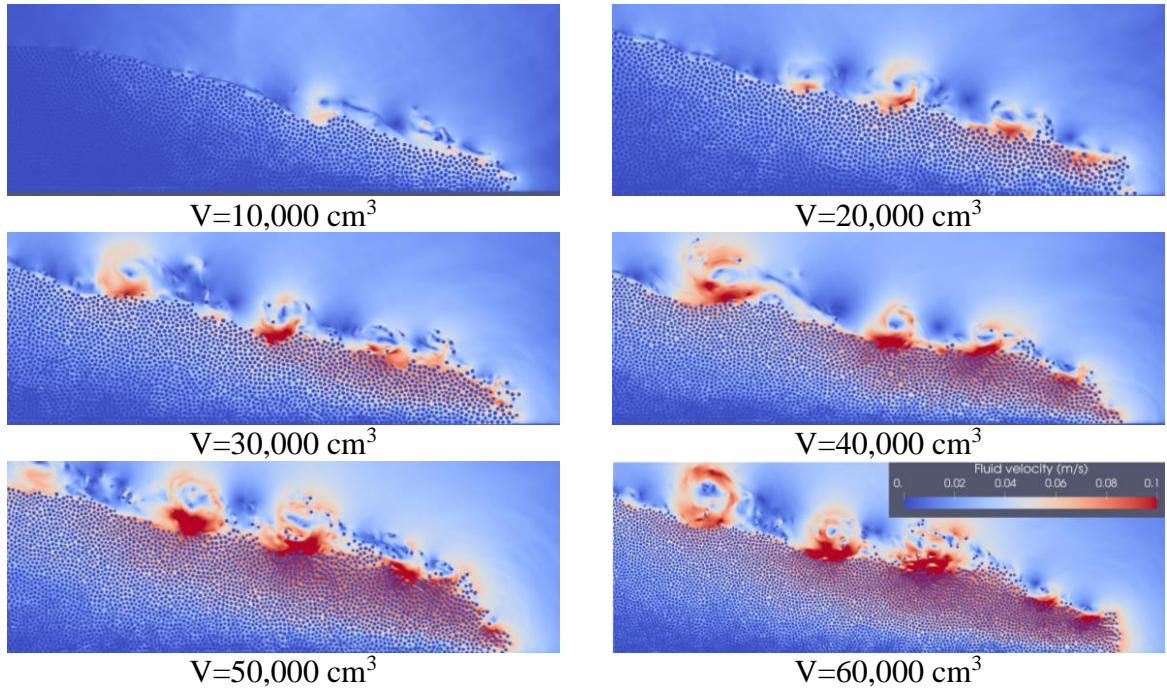

**Figure 8. Flow morphology of the flow front at $3\tau_c$ for different initial volumes**

**Table 1. The angle of attack under submerged and dry conditions at $3\tau_c$**

|  | Angle of attack (°) | | | |
|---|---|---|---|---|
|  | Submerged | | Dry | |
|  | Minimum | Maximum | Minimum | Maximum |
| V=10,000 cm³ | 39.5 | 51.2 | 30.7 | 40.8 |
| V=20,000 cm³ | 46.0 | 56.8 | 29.5 | 39.2 |
| V=30,000 cm³ | 44.0 | 57.5 | 29.9 | 40.3 |
| V=40,000 cm³ | 48.7 | 57.4 | 30.9 | 40.5 |
| V=50,000 cm³ | 52.9 | 61.4 | 30.1 | 41.5 |
| V=60,000 cm³ | 53.0 | 61.2 | 30.0 | 40.1 |

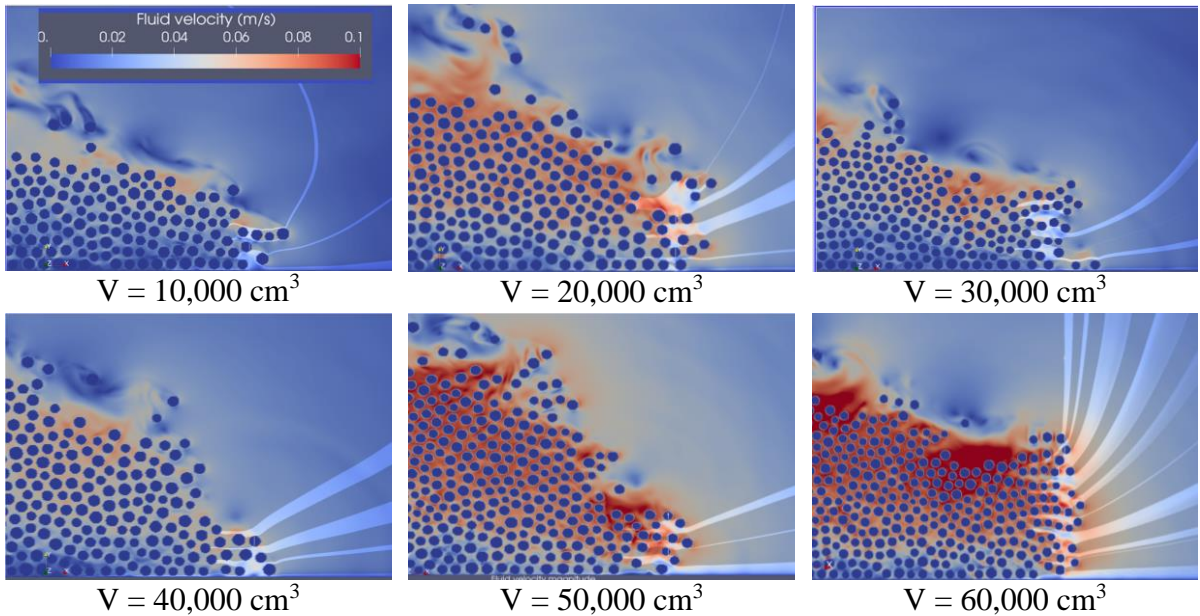

Figure 9. Streamlines at the flow front (~15 grain diameters) for different volumes at $3\tau_c$

**SUMMARY**

Two-dimensional LBM-DEM simulations were conducted to understand the effects of initial volume on the run-out behaviour of submerged granular column collapse. Six different initial volumes (V = 10,000, 20,000, 30,000, 40,000, 50,000 and 60,000 cm$^3$) of columns with an aspect ratio of 0.2 are simulated in both dry and submerged conditions. It is observed that: 1. In both dry and submerged cases, the final normalized run-out and the total normalized kinetic energy increase with volume. 2. The peak kinetic energy for submerged cases is 30-40% lower than the dry cases, and the total normalized kinetic energy is also lower for the submerged cases than the dry cases. 3. In the submerged cases, the run-out rate is lower, and the peak kinetic energy occurs later ($3\tau_c$ rather than $\tau_c$) compared to the dry cases. 4. For the submerged columns, the run-out rate and the normalized kinetic energy increases with volume at the initiation stage (t < 3 $\tau_c$). 5. For a volume of 60,000 cm$^3$ the run-out distance in the submerged case flows further that the dry case.

    The hydrodynamic mechanisms, such as negative pore pressure generation, hydroplaning, drag forces, and formation of turbulent vortices, affect the underwater run-out behaviour. Drag forces and turbulent vortices dissipate the kinetic energy of the granular flow. On the other hand, hydroplaning reduces the frictional dissipation at the base and inhibits energy loss. The influence of all these mechanisms increases with volume. For the smallest volume in this study, the effects of drag and vortices are more pronounced, resulting in a shorter final run-out distance than its dry counterparts. For the intermediate volumes, the inhibiting effects cancel out with the effect of hydroplaning, resulting in similar final run-out distances to their dry counterparts. For the largest volume, hydroplaning predominates, resulting in a larger run-out distance than its dry counterpart. Finally, slower run-out rates in the submerged cases compared to the dry cases are attributed to the generation of negative pore pressures in the initiation stage, the effect is more pronounced on smaller volumes.